# Review the Enterprise Resource Planning in Moroccan Healthcare Organizations

**Fatima Zahra Yamani[1], Mohamed El Merouani[2]**

[1] **Department of Science and Technology, Center of Doctoral Studies, Abdelmalek Essaadi University**
**Tetouan, 93000, Morocco**

[2] **Department of Science and Technology, Center of Doctoral Studies, Abdelmalek Essaadi University**
**Tetouan, 93000, Morocco**

**Abstract**

The Hospital Information Systems (HIS) in Morocco take a central place in the process of patient care. An approach is made to analyze the current situation of the HIS within the institutions in order to bring an integral and generic vision, allowing the judicious articulation of the business and IT layers. Currently, the Enterprise Resource Planning (ERP) implemented remains a system consisting of several applications dedicated to specific areas. These systems have become an indispensable element within any hospital. The goal of our study is to discover how the ERP has been used in Moroccan healthcare sector and how these software should be implemented and used to improve healthcare services.

## 1. Introduction

In recent years, there have been a series of important changes in the field of hospital information systems, healthcare institutions face lately the same constraints as production companies, for that, new approaches in this area go far beyond traditional approaches to find solutions for better management of the large volume of activities and for the massiveness of information and data. These establishments must comply with new management rules in order to manage the hospital effectively; among the important points that are part of permanent reform of health facilities, we mention the Enterprise Resource Planning which plays an important role in improving operational efficiencies, bettering communication and information flow among various departments.

The main goal of our work is to study the case of Moroccan health institutions, in order to measure the level of performance and efficiency of the ERP implemented. Our study will analyze then the existence ERP in the healthcare sector in Morocco and look for ways to improve healthcare services and ERPs implementation.

This paper is organized as follows: the second section presents the healthcare ERP, while the third defines its advantages and objectives. Business process and architecture will be discussed in the fourth section. The implementation of ERP in Moroccan organizations is the main topic of the fifth section, including the current problems, the challenges and the standards software requirements. The last section describes some concluding remarks and future work.

## 2. Healthcare ERP: History and definition

In 1990's, profound and important changes took place in the medical sector; organizations in general and health organizations in particular have become more and more interested about the development and the evolution of information system at different levels.

In 1960, Hospital Information System (HIS) was created; probably it is the first hospital information system. It was created in a hospital Kamynv, Mountain in 1971 [1], just after, many countries, including European countries have moved toward automation hospital information system since the early 1980, and until now, this system is still developing and evolving.

Enterprise Resource Planning is an important part of the organization's information system. It is considered the most groundbreaking progress in information technology as Al-Saleem and the other authors present on their analysis and overview of Enterprise Resource Planning [2]. EPR systems are software packages that allow an organization to view information about the entire organization, as well as the power to influence the operations of an organization [3].

In the domain of health, healthcare ERP is used to collect, store, process, readout and make communication between patients care with administrative data on all hospital activities and comply needs of all consumer systems. Its main goal is to support hospital activities at the practical, tactical and strategic levels [1].

In general, ERP systems in healthcare organizations include different software modules, which allow the hospital to automate and integrate the majority of functions by accessing and sharing common information, data and practices through the hospital in real-time [4].

## 3. Advantages and objectives

At the time, hospitals must be able to control its performance and information, and have the management tools particularly suited to its needs. They need automated information systems, such as an ERP system in order to meet the quality requirements of healthcare services.

ERP systems automate all organizational activities in a single IT system, improving access and sharing of organizational information and reducing repetitions and errors [5]. In addition, ERP systems help to reduce operational costs and streamline hospital processes. This helps the healthcare organization to manage and control various departments, since information is immediately available, to coordinate management and to ensure the efficiency of healthcare provided to patients [6].

The main advantages of ERP in healthcare are summarized below:

- Streamline healthcare processes
- Follow the care of patients,
- Provide users with highly secure access to patient's data
- Constitute all information, database concerning the patients and all the other related activities of the hospital,
- Every department and business unit of the organization will have access to readily available information when needed
- Carry out studies and medical research
- Traceability of the activities and acts
- Ease of knowledge sharing between stakeholders
- Reduce operational costs and increase profitability
- Improving staff efficiency
- Remove duplication and unnecessary procedures and use effective search facilities
- Improving quality of health care status.

## 4. Business Process and Architecture

The hospital is a complex and changing organization. It is considered to be a place of professionalization in which various professions evolve and cooperate, whose mission is the overall management of the patients' needs.

Figure 1 presents the different business processes in Moroccan healthcare organizations, from patient admission to discharge.

During patient care, the intervention of several parts in different departments is required, such as admission and discharge registered by admission service, transfer between departments, visit and determine the therapeutic actions orders issued by doctors and executed by nursing services, analysis realized by laboratory staff, perform diagnostic procedures and matters such as settlement and others. As we can observe, all these actions are performed separately for each patient and are carried out by different actors, for this, all the stakeholders need to access specific information about patients.

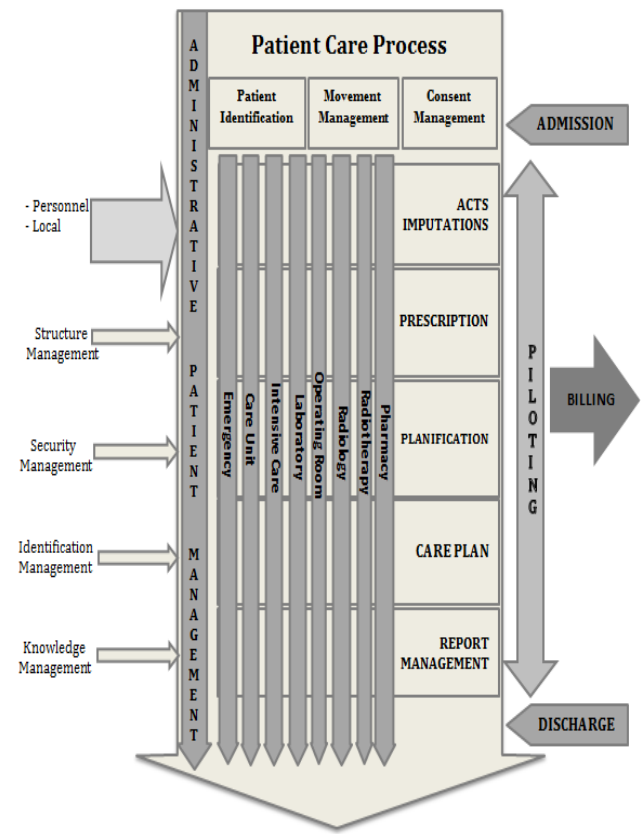

Fig. 1 Business Process in Moroccan Healthcare Organization

Figure 2 illustrates a central database of patients in a healthcare organization. Departments and trades mentioned above are represented as modules in the healthcare ERP.

These modules share the same database, which allows connectivity between integrated health services that share information for different purposes.

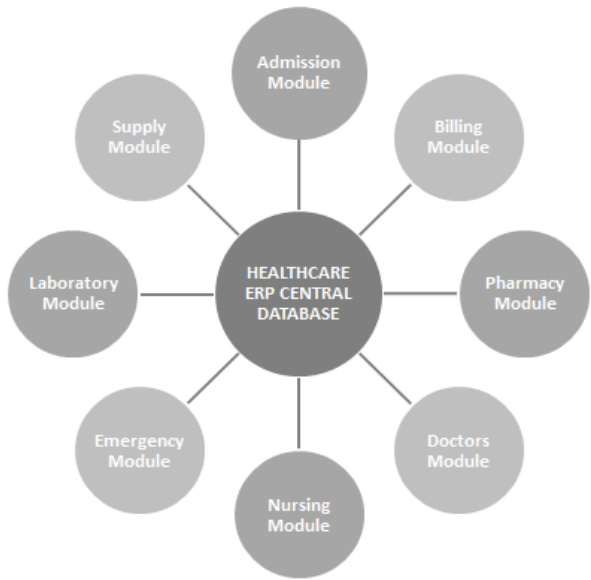

Fig. 2 Healthcare ERP Central Database

## 5. Healthcare ERP implementation in Moroccan organization

During the last 10 years, Morocco, which is a country in the process of development and whose implementation of the ERPs has become a strategic choice of multiple companies in different sectors of activity, turns to production companies to deploy an ERP system in healthcare organizations, these companies have solutions quite solid and adapted compared to the hospital sector.

The organization that is the object of our review is University Hospital Center (UHC). The first implementation of ERP system in Morocco was in UHC Mohammed VI of Marrakech in May, 2011, next, in 2020, the hospital Cheikh Zaid in Fes implements the same ERP, and later, in July, 2014, it was the start of the implementation in UHC Mohammed VI Oujda. The ERP implemented is a global solution made up of multiple applications, dedicated to several areas.

In order to evaluate the level of satisfaction, effectiveness and efficiency within these healthcare organizations, we are based on our work experience with one of these organizations, which gives us the opportunity to live problems and detect shortcomings in different departments and with different users.

Our main objective will be then studying the existing ERP, measuring the level of effectiveness, analyzing the current problems and challenges, then proposing some solutions.

### 5.1. Current problems

Although industrial field and hospital are similar on many points, they are different on many crucial elements. In the healthcare sector; we are not talking about products and machines, but about patients and doctors, that what makes hospital field very sensitive to errors and problems.

Actually, users of healthcare ERP face several problems which lead to the ineffectiveness and inefficiency of the implemented system. Firstly, the implementation of ERP is a complex task, it takes time to install, to make settings, to test and adapt the solution according to the need of users and departments. All these tasks generate a significant cost.

Once these tasks are completed, users should start using the tool; at this moment, they find difficulties to adapt to the new system. Users need simplicity and effectiveness in using ERP and ease of finding information, which does not exist in the current system. Furthermore, they claim the lack and insufficient training to easily use the tool.

On the other hand, the implemented ERP offers different subsystems. By being modular, each module can be used as a stand-alone solution, usually partitioned and functioning in a competing way, but just a problem on one of the subsystems, can affect other modules and block the work of other departments.

Moreover, due to the complexity and the evolution of healthcare organizations, the structure of the process and the number of entities [7] ERP is more and more evolutionary, which often requires the development of specific interfaces. The latter costs a lot and needs time to be applied, as well as it can affect other interfaces and cause problems after updates, especially when there is a lack of continuous and immediate maintenance.

Beside requests for development of specific interfaces, users signal the lack of interfaces that must have existed from the beginning without requiring development. All because of the requirements that have been poorly assessed and badly described.

In addition, the existing ERP are not all adapted to the volume of the establishment, to the Moroccan context, to the language used, to volume and characteristic of the Moroccan healthcare system.

### 5.2. Challenges

The implementation of Enterprise Resource Planning brings with it a set of challenges: The growth of the healthcare organizations necessarily involves a large volume of activities and therefore a large amount of information to manage. These organizations are largely impacted by the implementation of systems; the ERP must have a positive impact on organizations and improve the quality of the different processes.

On the other hand, Enterprise Resource Planning in healthcare becomes more complex and evolving, this is closely related to the complexity and scalability of healthcare organizations as well as the diversity of business processes [8].

Due to this complexity, there is often a lack of adequate analysis of requirements, which leads to the unavailability of some essential features and reduces productivity and profitability. Moreover, as ERP is in continuous evolution, it needs to be mastered and controlled, to better meet the needs of users at the right time.

We can also mention the lack of training and support; staff needs to have appropriate training during and after the implementation. It is very essential for them to be comfortable in using the software; this allows them to be more productive and satisfied with the use of the tool.

As well as support in case of errors or system bugs, or else, it will cause delay in operations and unnecessary frustrations in the organization.

One of the main challenges too, expensive ERP maintenance. This generates regressions and takes time. The risks of in-depth software modifications are avoided; it is preferable to add additional layers that ultimately increase the complexity [9]. Requirements should be discussed thoroughly and profoundly during the planning phase to minimize changes.

In addition, medical information in the hospital information system is more and more a major importance in a world where the increase in population requires the existence of an efficient information system [10] but in case of the incompatibility of the information system implemented with the needs of the organization, many problems will be faced. For this, the selection of the supplier of the ERP must be based on solid foundations.

### 5.3. Standards software requirements

This part presents some of the required standards for systems implementation and deployment solutions for the health sector in Morocco; we will focus more precisely on modeling.

As we have already mentioned in the previous parts, ERP healthcare is considered as one of the complex systems, especially when it touches a very sensitive area. For this, and as the author of [11] presents in his work, in order to understand a complex system, it is necessary to have a reading guide to understand the system in order to prepare a model close to reality and exploitable. We understand that modeling plays a crucial role in the development and efficiency of software; it is the representation of a real system in appropriate language, for the formalization and capitalization of knowledge in a form understandable and usable [12].

On the other hand, the author in [11] comments that there are many tools and methods for analyzing and modeling industrial systems, and even if each one provides solutions to several problems, none is sufficient for analyze and model complex systems. Hospital systems face the same problem, since they are complex and should be rich in features. For this, solutions in modeling must be proposed for the basic system, and that to have a minimum of problem on the software.

In short, to develop an ERP that responds to continuous improvement and changes in health organizations, we should build our system around trade and its interactions, not simply around databases and computer software. We should have a good analysis that is strongly related to the observation and modeling of patient's flow and the real system processes. With a good base of modeling, we will minimize changes in the interfaces, as well as it would be easy to add new interfaces if needed just by adding the necessary modifications to the models.

## 6. Conclusion and Future Work

Through this article, we presented four principal points, the definition of healthcare ERP, their advantages and objectives, the business process and architecture and finally their implementation in Moroccan organizations.

The paper allowed us to conclude several points; the first is that healthcare organizations are complex and sensitive to errors and problems, which makes the development of software as ERP a complex and difficult task. Secondly, the existing ERP implemented face several problems and challenges, some problems are related to the adaptation of users, lack of training and especially lack of functionalities. Finally, we noted that modeling is one of the important parts and it is the basis of good software.

In our future works, we will move towards an approach for developing healthcare ERP based on diagrams, our approach will improve the efficiency of software development, facilitate the frequent modifications in the code and minimize the significant effort to develop an efficient healthcare ERP.

**Fatima Zahra Yamani** got her Master Degree in IT management from Abdelmalek Essaadi University of Tetouan, Morocco in 2016. She is currently Ph.D student in Science and Technology with the research team of Modeling and Computer Theory at the center of doctoral studies in Tetouan, Morocco. Her research activities have focused on healthcare ERPs in Morocco using Model Driven Architecture approach.

**Mohamed El Merouani** obtained his Ph.D in Mathematics from University of Granada, Faculty of Sciences. He was a responsible of Department of Statistics and IT at the polydisciplinary faculty of Tetouan during many years. He is a professor and responsible of "Modeling and Computer Theory" research team. Currently, he is professor at the University Abdelmalek Essaadi of Tetouan (Morocco). He teaches several courses in the domain of mathematics and statistics.